\documentclass[notoc,paper,letterpaper]{JHEP3}
\usepackage{amssymb}
\def\be{\begin{equation}}
\def\ee{\end{equation}}
\def\bea{\begin{eqnarray}}
\def\eea{\end{eqnarray}}
\def\ie{{\it i.e.}}
\def\eg{{\it e.g.}}
\def\Z{\mathbb{Z}}
\def\vev#1{{\langle #1 \rangle}}
\def\Re{\mbox{Re}}
\def\a{{\alpha}}
\def\b{{\beta}}
\def\e{{\epsilon}}
\def\th{{\theta}}
\def\m{{\mu}}
\def\p{{\pi}}
\def\t{{\tau}}
\def\f{{\phi}}
\def\G{{\Gamma}}
\def\L{{\Lambda}}
\def\gg{{\bf g}}
\def\tf{{\bf t}}
\def\DD{{\cal D}}
\def\SS{{\cal S}}
\def\WW{{\cal W}}
\def\cN{{\cal N}}
\def\ra{{\rightarrow}}
\def\RR{{\mathbb{R}}}
\def\ZZ{{\mathbb{Z}}}
\def\ggL{{\gg^\vee}}
\def\Sv{{\tilde S}}

\title{On S-duality for Non-Simply-Laced Gauge Groups}

\author{Philip C. Argyres$^1$, Anton Kapustin$^2$, and
Nathan Seiberg$^3$\\
$^1$\,Physics Department, University of Cincinnati, 
Cincinnati OH 45221-0011
\vskip5pt 
$^2$\,California Institute of Technology, Pasadena, CA 91125
\vskip5pt
$^3$\,School of Natural Sciences, Institute for Advanced Study,
Princeton NJ 08540}

\abstract{We point out that for $\cN=4$ gauge theories with
exceptional gauge groups $G_2$ and $F_4$ the S-duality
transformation acts on the moduli space by a nontrivial
involution. We note that the duality groups of these theories are
the Hecke groups with elliptic elements of order six and four,
respectively. These groups extend the $\G_0(3)$ and $\G_0(2)$
subgroups of $SL(2,\Z)$ by elements with a non-trivial action on
the moduli space. We show that under a certain embedding of these
gauge theories into string theory, the Hecke duality groups are 
represented by T-duality transformations.}

\begin{document}

\section*{Introduction}

Strong-weak coupling duality, or S-duality, of $\cN=4$
super Yang-Mills (SYM) theory with gauge Lie algebra $\gg$ is the
conjectured \cite{mo77,o79} equivalence of this theory to a similar
theory with a magnetic-dual Lie algebra $\ggL$ and inverse gauge
coupling. (We recall the definition of $\ggL$
below; mathematicians refer to $\ggL$ as the Langlands-dual of $\gg$ 
and denote it ${}^L\gg$.) More precisely, let us define the complexified 
coupling $\t = (\th/2\p) + i(4\p/g^2)$, where
$g$ is the gauge coupling and $\theta$ is the theta-angle. For a 
simply-laced $\gg$, we have $\ggL=\gg$, 
and strong-weak coupling duality maps $\t$ to $-1/\t$. There is also 
a much more obvious symmetry $\t\ra\t +1$ which corresponds to shifting 
the theta-angle by $2\pi$. These two transformations together generate 
the group $SL(2,\Z)$ which acts on the coupling $\t$ by fractional linear
transformations: $\t \mapsto (a\t+b)/(c\t+d)$, where ${a\ b\choose
c\ d} \in SL(2,\Z)$.\footnote{The center of $SL(2,\ZZ)$ acts on the 
theory by charge-conjugation and leaves $\t$ invariant. If the Lie 
algebra does not have complex representations, then the duality group 
is $PSL(2,\ZZ)$ rather than $SL(2,\ZZ)$. This is the case for 
simply-laced Lie algebras ${\bf su}(2)$, $E_7,$ and $E_8$.}
The S-duality conjecture is supported by evidence
from the invariance of the effective action \cite{s9402a,dw9510,d0010}
and the BPS spectrum \cite{s9402b,ss95,p9505,dw9510,
dfhk9512,gl9601,lwy9601,g9603,dfhk9605,lll9606,fh9704,dhph9804} on the
moduli space of the SYM theories, as well as by the transformation
properties of the toroidally compactified partition function
\cite{ggpz9406,hms9501,ggpz9502} and of the 't Hooft-Wilson
operators \cite{k0501} in the conformal vacuum.  In addition,
related checks \cite{vw9408,bjsv9501} have been performed for
topologically twisted versions of $\cN=4$ SYM on more general
manifolds, which are sensitive to the gauge group, and not just
its Lie algebra.

For the non-simply-laced (compact simple) Lie algebras---$B_r$, $C_r$,
$G_2$ and $F_4$---the situation is more complicated.\footnote{We use
Dynkin notation for the simple Lie algebras: $A_r = {\bf su}(r+1)$, 
$B_r = {\bf so}(2r+1)$, $C_r = {\bf sp}(2r)$,
$D_r = {\bf so}(2r)$.}  The magnetic duals of these algebras are
$B_r^\vee=C_r$, $C_r^\vee=B_r$, $G_2^\vee=G_2'$, and $F_4^\vee=F_4'$,
where the primes on $G_2$ and $F_4$ indicate a rotation of their root
systems \cite{gno77} described below.  The studies \cite{ggpz9502,dw9510,
dfhk9605,dhph9804,k0501} of the partition function, BPS masses, and 't Hooft-Wilson
operators for general Lie algebras are consistent with the hypothesis
that strong-weak coupling duality maps the theories with Lie algebras 
$\gg$ and $\ggL$ to each other and acts by $\Sv:
\t\mapsto\t^\vee = -1/(q\tau)$ on the coupling.
Here $q$ is the ratio of the lengths-squared of long
and short roots of the Lie algebra $\gg$ or $\ggL$ (\ie, $q=2$ for $B_r$, 
$C_r$ and $F_4$, and $q=3$ for $G_2$).  When combined with the $2\pi$ 
shift of the theta-angle, 
the S-duality group acts on the coupling as an extension of
$\G_0(q)\subset SL(2,\Z)$ by the generator $\Sv$ \cite{dfhk9605}.
In the case $\gg=B_r$ or $C_r$, since $\Sv$ interchanges the
two Lie algebras, it is an equivalence between different theories,
and only the $\G_0(2)$ subgroup is the self-duality group.  For 
$\gg=G_2$ or $F_4$, however, the algebras are self-dual so $\Sv$ 
is supposed
to identify strongly-coupled $\gg$ with weakly-coupled $\gg$.
Furthermore, an argument using geometric engineering in type II
strings \cite{v9707} supports the conclusion that $G_2$ and $F_4$
are self-dual.  (An alternative is that there are new non-Lagrangian
$\cN=4$ theories which are the strong-coupling limits of the $G_2$ and
$F_4$ theories.)

The purpose of this note is to sharpen the statement about the action
of the conjectural S-duality groups for $G_2$ and $F_4$. As
pointed out in \cite{dfhk9605}, they are subgroups of $SL(2,\RR)$ not 
isomorphic to $SL(2,\ZZ)$.
These groups are known as Hecke groups.  We note that their actions on 
the electric and
magnetic charge lattices must involve a rotation that is not in
the Weyl group.  An implication of this is that these S-duality
groups not only act on the coupling and the electric and magnetic
charges, but also on the moduli space.  At self-dual values of the
coupling (fixed points of the action of the Hecke groups), this
means that certain discrete global symmetries are spontaneously
broken at generic points on the moduli space. We also show that
the unusual duality groups for $G_2$ and $F_4$ are realized as
T-duality groups in the string-theoretic approach of \cite{v9707}.

We briefly recall some definitions from the theory of Lie algebras;
see \eg\ \cite{h72} for an exposition.  At a generic point of the
moduli space, the gauge group is Higgsed to $U(1)^r\times \WW$, where
$\WW$ is the Weyl group of $\gg$.  This breaking is specified by picking
a Cartan subalgebra $\tf\subset\gg$.  The (unique up to rescaling)
Ad-invariant metric $\vev{\cdot,\cdot}$ on $\gg$ defines an isomorphism
between $\tf$ and its dual $\tf^*$ by $\vev{\a,\b}=\a^*(\b)$ for all
$\b\in\tf$.  The precise normalization of the metric will be fixed below.
We use this metric to identify $\tf$ and $\tf^*$ and henceforth
drop the $^*$'s.  The roots $\{\a\}$ of $\gg$---physically, the
$U(1)^r$ charges of the massive gauge bosons---span the root lattice
$\L_r$ in $\tf$.  The coroots $\{\a^\vee\}$ are then defined by
$\a^\vee := 2\a/\vev{\a,\a}$ and span the coroot lattice $\L_r^\vee$.
Physically, the coroots are magnetic charges of elementary BPS monopoles 
in the theory.
It follows from the structure of Lie algebras that the roots belong to
the dual of the coroot lattice and {\em vice versa}, that is,
$\vev{\a,\b^\vee}\in\Z$ for all roots $\a,\b$.  The dual of the root
lattice is the magnetic weight lattice, $\L_w^\vee$---the lattice of
magnetic charges allowed by the Dirac quantization condition.  Thus
$\L_w^\vee:=\L_r^*$ and $\L_r^\vee\subset\L_w^\vee$.  Likewise, the
electric charge lattice, or weight lattice, is $\L_w :=(\L_r^\vee)^*$
and $\L_r\subset\L_w$.  Any state or source is then labelled by its
electric and magnetic charges $(\e,\m)\in \L_w\oplus\L_w^\vee$.  The
Weyl group $\WW$ is a finite group generated by reflections $R_\a$
through the planes perpendicular to each root $\a$ which act on the
electric and magnetic charges by $R_\a: (\e,\m) \mapsto (\e{-}
\vev{\a^\vee,\e}\a , \m{-}\vev{\m,\a}\a^\vee)$.  Finally, magnetic dual
Lie algebras are defined as follows:  $\ggL$ is the magnetic dual of $\gg$
if its roots are the coroots of $\gg$.  So $(\ggL)^\vee=\gg$,
$\L_r(\ggL)=\L_r^\vee(\gg)$, and $\L_w(\ggL)=\L_w^\vee(\gg)$.
The list of magnetic dual Lie algebras is given in \cite{gno77}.

We also recall some facts related to the action of the $\G_0(q)$
subgroups of $SL(2,\Z)$ on the couplings and charges of $\cN=4$ SYM
theories \cite{ggpz9502,k0501}.  $SL(2,\Z)$ is generated by the three
elements $S={0\,-1\choose1\ \ 0}$, $T={1\ 1\choose0\ 1}$, $C=-1$, which
satisfy the relations $C^2 = 1$, $S^2 = C$, $(ST)^3 = C$, and $C$ is
central.  $\G_0(q)$ is the subgroup consisting of the matrices whose
lower left entry is a multiple of $q$.  It is generated by $C$,
$T$, and $S T^q S$.  $C$ is charge conjugation, which acts on charges
by $(\e,\m) \mapsto (-\e,-\m)$ and leaves the coupling constant
invariant.  Charge conjugation is a trivial operation for $F_4$ and 
$G_2$ because $-1$ belongs to the Weyl group.  If
we choose a normalization of the invariant metric on $\gg$ so that
short coroots have length $\sqrt2$, then the coefficient of the $\th$
parameter in the action is 1 for the minimal instanton, so that $\th$
is periodic with period $2\p$.  $T$ corresponds to the shift of $\th$
by $2\p$, and so acts by $\t\mapsto\t+1$ and $(\e,\m) \mapsto
(\e{+}\m,\m)$.  Finally, $ST^qS$ acts as $\t \mapsto \t/(1-q\t)$ and
$(\e,\m) \mapsto (-\e,q\e{-}\m)$.

We now examine $G_2$ and $F_4$ more closely.  Details of the root
systems of these algebras are tabulated in \cite{bmp85}, for example.

\section*{G$\bf{}_2$}

Though the Cartan subalgebra of $G_2$ is 2 dimensional, it is
convenient to describe $\tf$ as the plane orthogonal to $e_1+e_2+e_3$
in a 3-dimensional space with orthonormal basis $\{ e_i\}$, $i=1,2,3$.
Then the six short coroots of length $\sqrt2$ are $\pm(e_i-e_j)$ for
$i\neq j$, and the six long ones of length $\sqrt6$ are $\pm (2e_i -
e_j - e_k)$ for $i\neq j\neq k$.  It follows that the long roots are
the same as the short coroots, and the short roots are $1/3$ of the
long coroots.  Therefore, a transformation which takes the roots
to the coroots is $R^\vee: e_i \mapsto e_j-e_k$ for $(i,j,k) =
(1,2,3)$ and cyclic permutations. Upon restriction to the plane 
orthogonal to $e_1+e_2+e_3$, it becomes
a rotation by $\pi/2$ accompanied by a rescaling by a factor $\sqrt3$.

The Weyl group of $G_2$ is the dihedral group $\DD_6 \simeq \SS_3
\ltimes \Z_2$, where the $\SS_3$ acts by permutations of the $e_i$,
and the $\Z_2$ by $e_i\mapsto\pm e_i$. Elements of the Weyl group 
include rotations by
$\pi/3$ and reflections, but not $R^\vee/\sqrt3$.  Note, however, that
$(R^\vee)^2/3$, a rotation by $\p$, is an element of the Weyl group.

The moduli space is parametrized by the vacuum expectation values 
(VEVs) of six real scalars taking values in the Cartan subalgebra, 
which we write as $\f=\f_1 e_1+\f_2 e_2+\f_3 e_3$ with 
$\f_1+\f_2+\f_3=0$.\footnote{The scalars
transform as a vector of the $SO(6)_R$ symmetry.  For simplicity 
we consider only invariants made from a single component of this
vector.}  This space should be modded out by the Weyl group.  The 
adjoint Casimirs are a basis of Weyl invariant polynomials
in the $\f_i$'s.  They are clearly symmetric polynomials in the $\f_i^2$.
A basis is $s_2 := \sum_i \f_i^2$ and $s_6 := \prod_i \f_i^2$, where
the subscript on the $s_n$ denotes the scaling dimension.  (The dimension
four invariant $s_4 := \sum_{i<j} \f_i^2 \f_j^2$ is not independent
since the $\sum\f_i=0$ constraint implies $4s_4=s_2^2$.)  Note that
$s_2$ is determined up to a multiplicative factor by its scaling
dimension, while $s_6$ can be redefined by a multiplicative
factor as well as by the addition of a term proportional
to $s_2^3$.

The conjectural $\Sv$ transformation maps the coupling and charges
as $\t \mapsto -1/(3\t)$ and $(\e,\m) \mapsto (-R^\vee\m/3 , R^\vee\e)$.
Since $R^\vee/\sqrt3$ is not an element of the Weyl group, it will
have a non-trivial action on the moduli space. Indeed, the BPS mass 
formula
$$
M=\frac{|\f\cdot (\e+\m\t)|}{\sqrt {{\rm Im}\tau}}
$$
is invariant if in addition to transforming $\e, \m,$ and $\t$ as above 
one maps
$$
\Sv: \f\ra \frac{1}{\sqrt 3} R^\vee \f .
$$
Convenient coordinates on the moduli space are:
$$
U_2=s_2,\quad U_6:=s_6-(1/54)s_2^3.
$$
Then
\be\label{G2act}
\Sv: (U_2,U_6) \mapsto (U_2,-U_6),
\ee
These coordinates $U_2,U_6$ which transform homogeneously under $\Sv$ 
are unique up to overall multiplicative factors.

It is simple to see that the $ST^3S$ generator of $\G_0(3)$ is realized,
up to an overall rotation by the element $(R^\vee)^2/3$ of the Weyl
group, by $\Sv T\Sv$.  (Note that the other natural assignment
for the action of $\Sv$ on the charges, namely $(\e,\m) \mapsto
(-(R^\vee)^{-1}\m , R^\vee\e)$, fails to close on $\G_0(3)$.)  This
group, generated by $C$, $T$, and $\Sv$, is a type of Fuchsian
group known as a Hecke group \cite{b83}.  Its generators satisfy
the relations $C^2=1$, $\Sv^2=C$ and $(\Sv T)^6=C$ with $C$
central.\footnote{Actually, since $C$ is a trivial operation (it belongs
to the Weyl group), only the quotient of the Hecke subgroup by its center 
acts faithfully on the $G_2$ theory.}  A fundamental domain in the 
$\t$-plane is the region $|\t|
\ge 1/\sqrt3$ and $|\Re\,\t|\le1/2$ with boundaries identified so that
there is a $\Z_2$ orbifold point at $\t=i /\sqrt3$ and a $\Z_6$ orbifold
point at $\t=(i\pm\sqrt3)/(2\sqrt3)$. 

The $G_2$ $\cN=4$ SYM theory thus has an enhanced $\Z_2$ and $\Z_6$ global
symmetries at these special values of $\tau$.  However, at a generic point 
on the moduli space of vacua these symmetries are spontaneously broken as 
$\Z_2\to1$ and
$\Z_6\to\Z_3$, respectively, by virtue of the action (\ref{G2act}).

More generally, if all six Higgs fields are turned on, the moduli space 
is $\tf^{\otimes 6}$
modulo the diagonal action of the Weyl group. The transformation $\Sv$ 
acts as rotation by $\pi/2$
on all six copies of $\tf$. Once one identifies points on the moduli 
space related by the Weyl group,
the $\Sv$ transformation becomes an involution on the moduli space.

\section*{F$\bf{}_4$}

Take $\tf$ to be a four-dimensional space with an orthonormal basis
$\{e_i\}$, $i=1,2,3,4$.  The 24 short coroots of $F_4$ are
$\pm e_i \pm e_j$ (length $\sqrt2$) and the 24 long coroots are
$\pm 2e_i$ and $\pm e_1 \pm e_2 \pm e_3 \pm e_4$ (length $2$).
The long roots are the short coroots, while the short roots
are 1/2 the long coroots.  A transformation which takes the roots
to the coroots is $R^\vee: e_i \mapsto (R^\vee)^j_i e_j$ with
\be
R^\vee = \pmatrix{\phantom{-}1&1&&\cr -1&1&&\cr
&&\phantom{-}1&1\cr &&-1&1\cr},
\ee
a rotation by $\p/4$ in two orthogonal planes together with
a rescaling by a factor $\sqrt2$.

The Weyl group of $F_4$ is the group $\WW=\SS_3\ltimes(\SS_4\ltimes
\Z_2^3)$ where the $\SS_4$ in the second factor acts as permutations on
the $e_i$, $(\Z_2)^{3}$ acts as $e_i \mapsto \pm e_i$ with $\prod_i
(\pm)_i =1$, and the $\SS_3$ factor is generated by
\be
R_1 = {1\over2}\pmatrix{1&\phantom{-}1&\phantom{-}1&\phantom{-}1\cr
1&\phantom{-}1&-1&-1\cr 1&-1&\phantom{-}1&-1\cr 1&-1&-1&\phantom{-}1\cr}
\qquad\mbox{and}\qquad
R_2 = \pmatrix{-1&&&\cr &1&&\cr &&1&\cr &&&1\cr} .
\ee
Note that $R^\vee/\sqrt2 \not\in\WW$, but $(R^\vee)^2/2\in\WW$.

An adjoint scalar VEV on the moduli space can be parametrized by
$\f=\sum_i \f_i e_i$.  The Weyl invariant polynomials in the
$\f_i$'s are clearly symmetric polynomials in the $\f_i^2$ because
of the action of the $\SS_4\ltimes\Z_2^3$ factor of $\WW$ together
with the $\Z_2$ generated by $R_2$.  A basis of these polynomials
is $s_2 := \sum_i \f_i^2$, $s_4 := \sum_{i<j} \f_i^2 \f_j^2$, $s_6
:= \sum_{i<j<k} \f_i^2 \f_j^2 \f_k^2$, and $s_8 := \prod_i
\f_i^2$.  Combinations of these have to be further symmetrized
with respect to the $\SS_3$ factor generated by $R_1$ and $R_2$,
giving the four independent $\WW$-invariant polynomials
\bea\label{f4cas} U_2 &=& s_2, \qquad U_6 = 48 s_6 - 8 s_4 s_2 +
s_2^3, \qquad
U_8 = 48 s_8 - 6 s_6 s_2 + 4 s_4^2 - s_4 s_2^2 , \nonumber\\
U_{12} &=& 1152 s_8 s_4 - 360 s_8 s_2^2 - 216 s_6^2 + 72 s_6 s_4 s_2
- 12 s_6 s_2^3 - 32 s_4^3 + 12 s_4^2 s_2^2 - s_4 s_2^4 ,
\eea
whose dimensions are one plus the exponents of $F_4$.  Note that these
$U_n$ are determined up to multiplicative factors and addition of
appropriately homogeneous polynomials in the $U_m$ with $m<n$.  Such
additions could be used to simplify the above formulas for the $U_n$,
but the particular forms shown are chosen to have homogeneous
$R^\vee/\sqrt2$ transformation properties.

The conjectural $\Sv$ transformation maps coupling and charges as
$\t \mapsto -1/ (2\t)$ and $(\e,\m)\mapsto (-R^\vee\m/2,R^\vee\e)$.
A straightforward calculation then shows that it acts on the moduli 
space as
\be\label{F4act}
\Sv: (U_2,U_6,U_8,U_{12}) \mapsto (U_2,-U_6,U_8,-U_{12}).
\ee
Unlike the $G_2$ case, this homogeneous transformation law does not
completely determine the Casimirs up to overall rescalings, for $U_8$
may still be shifted by a multiple of $U_2^4$, and $U_{12}$
by a multiple of $U_2^3 U_6$.

The $ST^2S$ generator of $\G_0(2)$ is realized by $\Sv T\Sv$.
$C$, $T$, and $\Sv$ also generate a Hecke group, defined by the
relations $C^2=1$, $\Sv^2=C$, and $(\Sv T)^4=C$, with $C$ 
central.\footnote{Again, $C$ acts trivially, so it is the $\ZZ_2$ 
quotient of the Hecke group which acts faithfully on the $F_4$ theory.}
A fundamental domain in the $\t$-plane is the region $|\t|\ge 1/\sqrt2$
and $|\Re\,\t|\le1/2$ with boundaries identified so that there is
a $\Z_2$ orbifold point at $\t=i /\sqrt2$ and a $\Z_4$ orbifold point
at $\t=(i\pm1)/2$.  Thus the $F_4$ $\cN=4$ SYM theory has enhanced
$\Z_2$ and $\Z_4$ global symmetries at these values of the couplings, 
which are spontaneously broken to $1$ and $\Z_2$, respectively, on the
moduli space.

\section*{Stringy realization of the duality groups}

In \cite{v9707} it was shown how to embed $\cN=4$ SYM theory
with an arbitrary compact simple gauge Lie algebra into string theory 
so that S-duality
follows from a nontrivial symmetry of the worldsheet conformal
field theory (essentially, T-duality). This construction provides
an alternative way to derive the duality group for $G_2$ and $F_4$
theories.

To construct the $G_2$ theory, one starts with a six-dimensional
Little String Theory \cite{Seiberg:1997zk} obtained by taking a 
decoupling limit of Type
IIB string theory on a $D_4$ ALE singularity. Upon
compactification on a circle of radius $R_1$ this theory becomes
equivalent to a five-dimensional $\cN=2$ theory with gauge group
$SO(8)$. The coupling constant $1/g_5^2$ of this five-dimensional 
theory is
proportional to $R_1$. To obtain an $\cN=4$ theory in four
dimensions with gauge group $G_2$, one compactifies on a twisted
circle of radius $R_2$ \cite{v9707}. This means that one considers
an orbifold of the five-dimensional theory by a symmetry which
acts by the triality automorphism on the five-dimensional fields
combined with a translation of $x^5$ by $2\pi R_2$. Since the
triality-invariant part of the Lie algebra of $SO(8)$ is the Lie
algebra of $G_2$, this results in a four-dimensional $\cN=4$
theory with gauge group $G_2$. The coupling $1/g_4^2$ of this
theory is proportional to $R_1R_2$.

To fix the proportionality constant, we note that an instanton in
the 4d gauge theory is represented by a Euclidean fundamental string
worldsheet wrapping both circles. The action of such a worldsheet
instanton is
$$
2\pi \frac{R_1R_2}{\alpha'}.
$$
On the other hand, the action of an instanton in gauge theory is 
$-2\pi i\t$.  Hence we must identify
\begin{equation}\label{taueq}
\tau=i\frac{R_1R_2}{\alpha'}
\end{equation}
Since $\tau$ is purely imaginary, the theta-angle vanishes. To get
a nonzero theta-angle, one has to turn on the B-field flux,
resulting in
$$
\tau=\frac{i R_1 R_2 + B}{\alpha'}.
$$

It is shown in \cite{v9707} that for $B=0$ T-duality along both circles 
gives the same theory but with
$$R'_1=\frac{\alpha'}{R_1}, \quad R'_2=\frac{\alpha'}{3R_2}.$$
Then we have
$$
\tau'= i \frac{\alpha'}{3R_1R_2}=-\frac{1}{3\tau}.
$$
We also have a symmetry $\tau\ra\tau+1$ which corresponds to
shifting the B-field flux by $\alpha'$. These transformations
generate a Hecke subgroup of $SL(2,\RR)$, in agreement with the
field-theoretic approach.

The situation for $F_4$ is similar. One starts with a Little
String Theory obtained by considering the decoupling limit of Type
IIB string theory on a $E_6$ ALE singularity and compactifies it
on a circle of radius $R_1$. Then one orbifolds the resulting
five-dimensional theory by a transformation which acts by an outer
automorphism of $E_6$ of order $2$ and translates $x^5$ by $2\pi
R_2$. This gives a four-dimensional $\cN=4$ gauge theory with
gauge group $F_4$ \cite{v9707}. Its coupling is given by
(\ref{taueq}). It is shown in \cite{v9707} that T-duality maps
$$
R_1\mapsto R'_1=\frac{\alpha'}{R_1}, \quad R_2\mapsto
R'_2=\frac{\alpha'}{2R_2}.
$$
Hence $\tau'=-1/2\tau$, again in agreement with field theory.

\section*{Acknowledgments}

P.C.A. and A.K. would like to thank Institute for Advanced Study, 
Princeton, NJ, for hospitality while this work was being performed.
P.C.A is supported in part by DOE grant DOE-FG02-84-ER40153.
A.K. is supported in part by DOE grant DOE-FG03-92-ER40701.
N.S. is supported in part by DOE grant DOE-FG02-90-ER40542.

\end{document}